\documentclass[preprint,aps,showpacs]{revtex4}
\usepackage{amsmath}
\usepackage{ulem}
\usepackage{bm}
\usepackage{graphics}
\usepackage[dvips]{graphicx,color}
\usepackage{dcolumn}

\begin{document}
\title{Triplet pair amplitude in a trapped $s$-wave superfluid Fermi gas with broken spin rotation symmetry}
\author{Yuki Endo, Daisuke Inotani, Ryo Hanai, and Yoji Ohashi}
\affiliation{Department of Physics, Faculty of Science and Technology, Keio University, 3-14-1, Hiyoshi, Kohoku-ku, Yokohama 223-8522, Japan}
\date{\today}
\begin{abstract}
We investigate the possibility that the broken spatial inversion symmetry by a trap potential induces a spin-triplet Cooper-pair amplitude in an $s$-wave superfluid Fermi gas. Being based on symmetry considerations, we clarify that this phenomenon may occur, when a spin rotation symmetry of the system is also broken. We also numerically confirm that a triplet pair amplitude is really induced under this condition, using a simple model. Our results imply that this phenomenon is already present in a trapped $s$-wave superfluid Fermi gas with spin imbalance. As an interesting application of this phenomenon, we point out that one may produce a $p$-wave superfluid Fermi gas, by suddenly changing the $s$-wave pairing interaction to a $p$-wave one by using the Feshbach resonance technique. Since a Cooper pair is usually classified into the spin-singlet (and even-parity) state and the spin-triplet (and odd-parity) state, our results would be useful in considering how to mix them with each other in a superfluid Fermi gas. Such admixture has recently attracted much attention in the field of non-centrosymmetric superconductivity, so that our results would also contribute to the further development of this research field, on the viewpoint of cold Fermi gas physics.
\end{abstract}
\pacs{03.75.Ss, 03.75.-b, 67.85.Lm}
\maketitle
\par
\section{Introduction}
\par
Since any experiment on a superfluid Fermi gas is done in trap potential\cite{Jin,Zwierlein,Kinast,Bartenstein,Ketterle,Hulet,Shin}, it is interesting to explore physical phenomena originating from this spatial inhomogeneity. An example is surface oscillations observed in a $^6$Li superfluid Fermi gas\cite{Kinast,Bartenstein}. Another example is the phase separation observed in a $^6$Li Fermi gas with spin imbalance\cite{Ketterle,Hulet,Shin}, where the spin-balanced superfluid region in the trap center is spatially surrounded by excess atoms. 
\par
In addition to these macroscopic phenomena, the spatial inhomogeneity can also affect microscopic superfluid properties. Noting that a trap potential breaks the spatial inversion symmetry when the inversion center is taken to be away from the trap center, we expect that the parity becomes no longer a good quantity to classify the spatial structure of a Cooper pair, leading to the admixture of even and odd parity symmetry. Since a pair wavefunction is always antisymmetric with respect to the exchange of two fermions, this naturally leads to the mixing of spin-singlet and spin-triplet state. When this phenomenon occurs, the $s$-wave superfluid state is accompanied by a triplet Cooper pair amplitude, in addition to the ordinary singlet component. (The Cooper pair amplitude is symbolically written as $\langle c_{{\bm p},\alpha} c_{-{\bm p},\alpha'}\rangle$, where $c_{{\bm p},\alpha}$ is an annihilation operator of a Fermi atom with pseudospin $\alpha=\uparrow,\downarrow$.)
\par
The purpose of this paper is to theoretically explore this possibility in a trapped $s$-wave superfluid Fermi gas. Using symmetry considerations, we prove that this phenomenon may occur, when a spin rotation symmetry of this system is also broken, in addition to the broken inversion symmetry by a trap potential. In a two-component Fermi gas, this additional condition is realized, when two species feel different trap potentials or chemical potentials, or when they have different atomic masses. Although this is a necessary condition, we numerically confirm that a triplet pair amplitude is really induced under this condition, within the mean-field theory for a model two-dimensional lattice Fermi superfluid in a harmonic trap.
\par
In considering a triplet pair amplitude, one should note that the appearance of this quantity does not immediately mean the realization of a triplet superfluid state. Actually, the system is still in the $s$-wave superfluid state, as far as the system only has an $s$-wave interaction. This is simply because the symmetry of a Fermi superfluid is fully determined by the symmetry of the superfluid order parameter, which is essentially given by the product of a pairing interaction and a pair amplitude. For example, an $s$-wave superfluid Fermi gas with a contact type $s$-wave pairing interaction $-U_s$ ($<0)$ is characterized by the ordinary $s$-wave superfluid order parameter,
\begin{equation}
\Delta_s=U_s\sum_{\bm p}
\langle
c_{{\bm p},\uparrow} c_{-{\bm p},\downarrow}
\rangle,
\label{eq.1}
\end{equation}
which is finite when the pair amplitude $\langle c_{{\bm p},\uparrow} c_{-{\bm p},\downarrow}\rangle$ has the $s$-wave component. The odd-parity component does not contribute to $\Delta_s$ in Eq. (\ref{eq.1}). 
\par
However, for an $s$-wave superfluid Fermi gas with both the singlet and triplet pair amplitude, when one suddenly changes the $s$-wave pairing interaction to a triplet (and odd parity) one $U({\bm p},{\bm p}')$, while the $s$-wave superfluid order parameter in Eq. (\ref{eq.1}) immediately vanishes due to the vanishing $s$-wave interaction ($U_s=0$), the product of the triplet interaction and the triplet component in the pair amplitude  $\langle c_{{\bm p},\uparrow} c_{-{\bm p},\downarrow}\rangle$ (which is assumed to have already existed in the $s$-wave state) immediately gives a finite triplet superfluid order parameter,
\begin{equation}
\Delta({\bm p})=\sum_{{\bm p}'}U({\bm p},{\bm p}')
\langle
c_{{\bm p}',\uparrow} c_{-{\bm p}',\downarrow}
\rangle,
\label{eq.2}
\end{equation}
when the triplet interaction $U({\bm p},{\bm p}')$ is chosen so that the momentum summation in Eq. (\ref{eq.2}) can be finite. In an ultracold Fermi gas, the change of the interaction is possible by using a tunable interaction associated with a Feshbach resonance\cite{Chin,Regal,Ticknor,Zhang,Nakasuji,Gaebler2007,Inada2008,Fuchs2008}. Then, by definition, one obtains a triplet superfluid Fermi gas characterized by the superfluid order parameter $\Delta({\bm p})$ in Eq. (\ref{eq.2}), at least just after this manipulation. This makes us expect that, when one can induce a $p$-wave pair amplitude in an $s$-wave superfluid Fermi gas, a $p$-wave superfluid Fermi gas may be realized. This possibility has recently been discussed by one of the authors\cite{Yamaguchi}, where a $p$-wave pair amplitude is induced by a synthetic spin-orbit interaction\cite{SOC0,Rb-SOC1,Rb-SOC2,Rb-SOC3,Rb-SOC4,Li-SOC,K-SOC,coldSOC5,coldSOC6}. The present paper provides another source of $p$-wave pair amplitude, without using an artificial gauge field.
\par
The admixture of singlet and triplet Cooper pairs has recently attracted much attention in the field of non-centrosymmetric superconductivity\cite{Bauer2004,Sigrist}, where a crystal lattice with no inversion center causes this phenomenon. In this field, it has been pointed out that this admixture may be the origin of the anomalous temperature dependence of the penetration depth observed in Li$_2$Pt$_3$B\cite{Yuan}. Thus, an $s$-wave superfluid Fermi gas with a triplet pair amplitude would be also helpful to the study of this electron system. 
\par
This paper is organized as follows. In Sec.II, we clarify the necessary condition for a triplet Cooper pair amplitude to appear in a trapped $s$-wave superfluid Fermi gas. In Sec.III, we numerically evaluate how large a triplet pair amplitude is induced under the condition obtained in Sec.II. In this section, we treat a superfluid Fermi gas loaded on a two-dimensional square lattice, within the mean-field theory. Throughout this paper, we take $\hbar=k_{\rm B}=1$, for simplicity.
\par
\par
\par
\section{Condition for triplet pair amplitude to appear in a trapped $s$-wave superfluid Fermi gas}
\par
We consider a three-dimensional $s$-wave superfluid Fermi gas, described by the Hamiltonian,
\begin{equation}
H=\int d{\bm r}
\Bigl[
\sum_\alpha
\Psi^\dagger_\alpha({\bm r})
h_{\alpha,\alpha'}({\bm r})
\Psi_{\alpha'}({\bm r})
-U_s
\Psi^\dagger_\uparrow({\bm r})
\Psi^\dagger_\downarrow({\bm r})
\Psi_\downarrow({\bm r})
\Psi_\uparrow({\bm r})
\Bigr].
\label{eq.3}
\end{equation}
Here, $\Psi_\alpha({\bm r})$ is a fermion field operator with pseudospin $\alpha=\uparrow,\downarrow$, describing two atomic hyperfine states. $-U_s$ $(<0)$ is a contact-type $s$-wave pairing interaction. $h_{\alpha,\alpha'}({\bm r})$ is a one-particle Hamiltonian density, consisting of a kinetic term and a potential term, detailed expression of which will be given later. 
\par
We assume that the system is in the ordinary $s$-wave superfluid state with the $s$-wave superfluid order parameter, 
\begin{equation}
\Delta_s({\bm r})=U_s\langle\Psi_\uparrow({\bm r})\Psi_\downarrow({\bm r})\rangle.
\label{eq.4}
\end{equation}
We also assume that any other spontaneous symmetry breaking is absent (such as the triplet superfluid state).
\par
In this model superfluid, we consider the spin-triplet Cooper-pair amplitude, given by
\begin{eqnarray}
\Phi_{\rm t}^{S_z}({\bm r},{\bm r}')=
\left\{
\begin{array}{ll}
\langle\Psi_\uparrow({\bm r})\Psi_\uparrow({\bm r}')\rangle
&
~~(S_z=1),\\
{1 \over \sqrt{2}}
\left[
\langle\Psi_\uparrow({\bm r})\Psi_\downarrow({\bm r}')\rangle
+
\langle\Psi_\downarrow({\bm r})\Psi_\uparrow({\bm r}')\rangle
\right]
&
~~(S_z=0),
\\
\langle\Psi_\downarrow({\bm r})\Psi_\downarrow({\bm r}')\rangle
&
~~(S_z=-1),
\end{array}
\right.
\label{eq.6}
\end{eqnarray}
where $S_z$ denotes the $z$-component of the total spin of each pair amplitude. The triplet pair amplitude in Eq. (\ref{eq.6}) does not contribute the $s$-wave superfluid order parameter $\Delta_s({\bm r})$ in Eq. (\ref{eq.4}), because $\Phi_{\rm t}^{S_z}({\bm r},{\bm r})=0$. The spin-singlet pair amplitude, 
\begin{equation}
\Phi_{\rm s}({\bm r},{\bm r}')=
{1 \over \sqrt{2}}
\left[
\langle\Psi_\uparrow({\bm r})\Psi_\downarrow({\bm r}')\rangle
-
\langle\Psi_\downarrow({\bm r})\Psi_\uparrow({\bm r}')\rangle
\right],
\label{eq.5}
\end{equation}
only contributes to Eq. (\ref{eq.4}).
\par
We first prove that the broken spatial inversion symmetry is necessary for a triplet pair amplitude to appear in an $s$-wave superfluid Fermi gas. For this purpose, we conveniently introduce the inversion operator ${\hat P}({\bm R})$ with respect to the inversion center ${\bm R}$. The field operator is transformed under this operation as
\begin{equation}
{\tilde \Psi}_\alpha({\bm r})
\equiv {\hat P}({\bm R})\Psi_\alpha({\bm r}){\hat P}^{-1}({\bm R})=
\Psi_\alpha({\bm R}-{\bm l}),
\label{eq.7}
\end{equation}
where ${\bm r}={\bm R}+{\bm l}$. The inverted Hamiltonian ${\tilde H}={\hat P}H{\hat P}^{-1}$ is then given by
\begin{eqnarray}
{\tilde H}
&=&\int d{\bm l}
\Bigl[
\sum_\alpha
\Psi^\dagger_\alpha({\bm R}-{\bm l})
h_{\alpha,\alpha'}({\bm R}+{\bm l})
\Psi_{\alpha'}({\bm R}-{\bm l})
\nonumber
\\
&-&U_s
\Psi^\dagger_\uparrow({\bm R}-{\bm l})
\Psi^\dagger_\downarrow({\bm R}-{\bm l})
\Psi_\downarrow({\bm R}-{\bm l})
\Psi_\uparrow({\bm R}-{\bm l})
\Bigr].
\label{eq.8}
\end{eqnarray}
When the one-particle Hamiltonian density $h_{\alpha,\alpha'}({\bm r})$ has the  symmetry $h_{\alpha,\alpha'}({\bm R}+{\bm l})=h_{\alpha,\alpha'}({\bm R}-{\bm l})$, this system is invariant (${\tilde H}=H$) under this symmetry operation. On the other hand, the triplet pair amplitude $\Phi_{\rm t}^{S_z=1}({\bm r},{\bm r}')$ in Eq. (\ref{eq.6}) with the center of mass position ${\bm R}=[{\bm r}+{\bm r}']/2$ is transformed as,
\begin{eqnarray}
{\tilde \Phi}_{\rm t}^{S_z=1}({\bm r},{\bm r}')
&\equiv&
\langle
{\tilde \Psi}_\uparrow({\bm r}){\tilde \Psi}_\uparrow({\bm r}')
\rangle
=
\langle
\Psi_\uparrow({\bm R}-{\bm r}_{\rm rel}/2)\Psi_\uparrow({\bm R}+{\bm r}_{\rm rel}/2)
\rangle
\nonumber
\\
&=&
-\langle
\Psi_\uparrow({\bm R}+{\bm r}_{\rm rel}/2)\Psi_\uparrow({\bm R}-{\bm r}_{\rm rel}/2)
\rangle
=-\Phi_{\rm t}^{S_z=1}({\bm r},{\bm r}'),
\label{eq.9}
\end{eqnarray}
where ${\bm r}_{\rm rel}={\bm r}-{\bm r}'$ is the relative coordinate. We also find ${\tilde \Phi}_{\rm t}^{S_z=0,-1}({\bm r},{\bm r}')=-\Phi_{\rm t}^{S_z=0,-1}({\bm r},{\bm r}')$. That is, the triplet pair amplitude $\Phi_{\rm t}^{S_z}({\bm r},{\bm r}')$ vanishes, when the system has the inversion symmetry (${\tilde H}=H$) with respect to the inversion center ${\bm R}=[{\bm r}+{\bm r}']/2$. Thus, the broken inversion symmetry is necessary for a triplet pair amplitude to appear. 
\par
For the singlet pair amplitude in Eq. (\ref{eq.5}), this symmetry operation simply gives ${\tilde \Phi}_{\rm s}({\bm r},{\bm r}')=\Phi_{\rm s}({\bm r},{\bm r}')$. As expected, this quantity may be finite.
\par
The one-particle Hamiltonian density $h_{\alpha,\alpha'}({\bm r})$ in the ordinary uniform Fermi gas has the form
\begin{equation}
h_{\alpha,\alpha'}({\bm r})=
\Bigl[
{{\hat {\bm p}}^2 \over 2m}-\mu
\Bigr]\delta_{\alpha,\alpha'},
\label{eq.10}
\end{equation}
where ${\hat {\bm p}}=-i\nabla$, $m$ is an atomic mass, and $\mu$ is the Fermi chemical potential. Equation (\ref{eq.10}) has the symmetry property, $h_{\alpha,\alpha'}({\bm R}+{\bm l})=h_{\alpha,\alpha'}({\bm R}-{\bm l})$, with respect to ${\bm l}$ for an arbitrary ${\bm R}$. To conclude, any triplet pair amplitude is not induced. 
\par
In the presence of a harmonic trap, the one-particle Hamiltonian density becomes inhomogeneous as
\begin{equation}
h_{\alpha,\alpha'}({\bm r})=
\Bigl[
{{\hat {\bm p}}^2 \over 2m}-\mu+{1 \over 2}Kr^2
\Bigr]
\delta_{\alpha,\alpha'},
\label{eq.11}
\end{equation}
so that it does not have the inversion symmetry except at ${\bm R}=0$. However, when we consider the $s$-wave superfluid state in this trapped case, any triplet pair amplitude is not actually induced (although we do not explicitly show the result here). Of course, since the condition obtained from the inversion symmetry is a {\it necessary} condition, the broken inversion symmetry does not guarantee the appearance of a triplet pair amplitude. 
\par
In this regard, we point out that the vanishing triplet pair amplitude in the trapped case is due to the fact that this system still has a rotation symmetry in spin space. To see this, we next consider the spin rotation of the field operator, given by
\begin{equation}
{\tilde \Psi}_\alpha({\bm r})=
{\hat R}({\bm \theta})\Psi_{\alpha}({\bm r}){\hat R}^{-1}({\bm \theta})=
\sum_{\alpha'}
\Bigl(
e^{{i \over 2}{\bm \theta}\cdot{\bm \sigma}}
\Bigr)_{\alpha,\alpha'}\Psi_{\alpha'}({\bm r}).
\label{eq.12}
\end{equation}
Here, ${\bm \theta}=\theta {\bm e}_\theta$ describes the spin rotation around the unit vector ${\bm e}_\theta$ with the angle $\theta$, and ${\bm \sigma}=(\sigma_x,\sigma_y,\sigma_z)$, where $\sigma_j$ ($j=x,y,z)$ are Pauli matrices. (As usual, we take the spin quantization axis in the $\sigma_z$-direction.) For the three ``$\pi$ rotations" specified by ${\bm \theta}=(\pi,0,0)~(\equiv{\bm \theta}_\pi^x)$, $(0,\pi,0)~(\equiv{\bm \theta}_\pi^y)$, $(0,0,\pi)~(\equiv{\bm \theta}_\pi^z)$, Eq. (\ref{eq.12}) can be written as,
\begin{equation}
\left(
\begin{array}{c}
{\tilde \Psi}_\uparrow({\bm r}) \\
{\tilde \Psi}_\downarrow({\bm r})
\end{array}
\right)_{{\bm \theta}={\bm \theta}_\pi^j}
=
i\sigma_j
\left(
\begin{array}{c}
\Psi_\uparrow({\bm r}) \\
\Psi_\downarrow({\bm r})
\end{array}
\right),
\label{eq.13}
\end{equation}
where we have used the formula $e^{i\frac{\theta}{2}\sigma_j}=\cos(\theta/2)+i\sigma_j\sin(\theta/2)$. Under the $\pi$-rotation, the Hamiltonian in Eq. (\ref{eq.3}) is transformed as 
\begin{eqnarray}
{\tilde H}
&=&
{\hat R}({\bm \theta}_\pi^j) H{\hat R}^{-1}({\bm \theta}_\pi^j)
\nonumber
\\
&=&
\int d{\bm r}
\Bigl[
\sum_\alpha
{\tilde \Psi}^\dagger_\alpha({\bm r})
h_{\alpha,\alpha'}({\bm r})
{\tilde \Psi}_{\alpha'}({\bm r})
-U_s
{\tilde \Psi}^\dagger_\uparrow({\bm r})
{\tilde \Psi}^\dagger_\downarrow({\bm r})
{\tilde \Psi}_\downarrow({\bm r})
{\tilde \Psi}_\uparrow({\bm r})
\Bigr]
\nonumber
\\
&=&
\int d{\bm r}
\Bigl[
\sum_\alpha
\Psi^\dagger_\alpha({\bm r})
\left(
\sigma_j{\hat h}({\bm r})\sigma_j
\right)_{\alpha,\alpha'}
\Psi_{\alpha'}({\bm r})
-U_s
\Psi^\dagger_\uparrow({\bm r})
\Psi^\dagger_\downarrow({\bm r})
\Psi_\downarrow({\bm r})
\Psi_\uparrow({\bm r})
\Bigr].
\label{eq.14}
\end{eqnarray}
Here, ${\hat h}({\bm r})=\{h_{\alpha,\alpha'}({\bm r})\}$. Thus, one has ${\tilde H}=H$, when
\begin{equation}
[{\hat h}({\bm r}),\sigma_j]=0
\label{eq.15}
\end{equation}
is satisfied. 
\par
While the singlet pair amplitude in Eq. (\ref{eq.5}) remains unchanged under these $\pi$-rotations, the triplet component is transformed as
\begin{eqnarray}
\left(
\begin{array}{c}
{\tilde \Phi}_{\rm t}^{S_z=1}({\bm r},{\bm r}') \\
{\tilde \Phi}_{\rm t}^{S_z=0}({\bm r},{\bm r}') \\
{\tilde \Phi}_{\rm t}^{S_z=-1}({\bm r},{\bm r}')
\end{array}
\right)
=
\left\{
\begin{array}{ll}
\left(
\begin{array}{c}
-\Phi_{\rm t}^{S_z=-1}({\bm r},{\bm r}') \\
-\Phi_{\rm t}^{S_z=0}({\bm r},{\bm r}') \\
-\Phi_{\rm t}^{S_z=1}({\bm r},{\bm r}')
\end{array}
\right)& ({\bm \theta}={\bm \theta}_\pi^x),\\
\left(
\begin{array}{c}
\Phi_{\rm t}^{S_z=-1}({\bm r},{\bm r}') \\
-\Phi_{\rm t}^{S_z=0}({\bm r},{\bm r}') \\
\Phi_{\rm t}^{S_z=1}({\bm r},{\bm r}')
\end{array}
\right)& ({\bm \theta}={\bm \theta}_\pi^y),\\
\left(
\begin{array}{c}
-\Phi_{\rm t}^{S_z=1}({\bm r},{\bm r}') \\
\Phi_{\rm t}^{S_z=0}({\bm r},{\bm r}') \\
-\Phi_{\rm t}^{S_z=-1}({\bm r},{\bm r}')
\end{array}
\right)& ({\bm \theta}={\bm \theta}_\pi^z).\\
\end{array}
\right.
\label{eq.16}
\end{eqnarray}
For example, when we set ${\bm \theta}={\bm \theta}_\pi^x$, Eq. (\ref{eq.16}) means that $\Phi_{\rm t}^{S_z=0}({\bm r},{\bm r}')=0$, when $[{\hat h}({\bm r}),\sigma_x]=0$. (The other two components with $S_z=\pm 1$ are not excluded in this case.) When the system invariant under all the $\pi$-rotations (${\bm \theta}_\pi^{x,y,z}$), any triplet pair amplitude is not induced. 
\par
To conclude, the broken spin rotation symmetry characterized by ${\bm \theta}_\pi^j$ is necessary for a triplet pair amplitude to be induced in a trapped $s$-wave superfluid Fermi gas. This is the reason why the model case described by Eqs. (\ref{eq.3}) and (\ref{eq.11}) is not accompanied by any triplet pair amplitude.
\par
The broken spin rotation symmetry is realized, when the strength of trap potential $K$ in Eq. (\ref{eq.11}) depends on spin ($\equiv K_\alpha$). In this case, the one-particle Hamiltonian ${\hat h}({\bm r})=\{h_{\alpha,\alpha'}({\bm r})\}$ can be written as
\begin{equation}
{\hat h}({\bm r})
=
\Bigl[
{{\hat {\bm p}}^2 \over 2m}-\mu+{K_\uparrow+K_\downarrow \over 4}r^2
\Bigr]
+{K_\uparrow-K_\downarrow \over 4}r^2\sigma_z.
\label{eq.17}
\end{equation}
Equation (\ref{eq.15}) is then satisfied only when $j=z$. Thus, we find from the last line in Eq. (\ref{eq.16}) that only the triplet pair amplitude with $S_z=0$ may be induced. Since the last term in Eq. (\ref{eq.17}) works as an external magnetic field, this phenomenon is also expected in the presence of spin imbalance\cite{Ketterle,Hulet,Shin,Nascimbene,Kashimura}, where two species feel different Fermi chemical potentials $\mu_\uparrow\ne\mu_\downarrow$. Another possibility is a trapped hetero superfluid Fermi gas\cite{Taglieber2008,Wille2008,Spiegelhalder2009,Takemori2012,Lin2006,Dao2007,Cazalilla2005,Hanai2013,Hanai2014}, where two species have different atomic masses $m_\uparrow\ne m_\downarrow$. In Sec. III, we will numerically examine these cases.
\par
Before ending this section, we briefly note that the broken inversion symmetry, as well as the broken spin rotation symmetry, are also realized in a spin-orbit coupled $s$-wave superfluid Fermi gas\cite{SOC0,Rb-SOC1,Rb-SOC2,Rb-SOC3,Rb-SOC4,Li-SOC,K-SOC,coldSOC5,coldSOC6}. Indeed, Refs.\cite{Yamaguchi,Hu2011,2B1} predict that a $p$-wave pair amplitude is induced in this case. Although we do not deal with this case in Sec.III, we explain in Appendix A how to apply the present symmetry consideration to this case.
\par
\par
\section{Numerical confirmation for the induction of triplet pair amplitude in a trapped $s$-wave superfluid Fermi gas}
\par
To examine whether or not a triplet pair amplitude is induced under the condition obtained in Sec. II, we consider a model $s$-wave Fermi superfluid loaded on a $L\times L$ two-dimensional square lattice, within the mean-field approximation. Although this simple model cannot be directly applied to a real continuum Fermi gas, it is still helpful to grasp basic characters of this phenomenon.
\par
The Hamiltonian is given by
\begin{eqnarray}
H_{\rm MF}
&=&-\sum_{\langle i,j\rangle,\alpha}t_\alpha
\Bigl[
c_{{\bm r}_i,\alpha}^\dagger c_{{\bm r}_j,\alpha}+{\rm h.c.}
\Bigr]
+\sum_i\Delta_s({\bm r}_i)
\Bigl[
c^\dagger_{{\bm r}_i,\uparrow}c^\dagger_{{\bm r}_i,\downarrow}+{\rm h.c.}
\Bigr]
\nonumber
\\
&+&\sum_{i,\alpha}
\Bigl[
V_\alpha({\bm r}_i)-\mu_{\alpha}-U_s n_{-\alpha}({\bm r}_i)
\Bigr]c_{{\bm r}_i,\alpha}^\dagger c_{{\bm r}_i,\alpha}.
\label{eq.18}
\end{eqnarray}
Here, $c^\dagger_{{\bm r}_i,\alpha}$ is a creation operator of a Fermi atom at the lattice site ${\bm r}_i=(r_x^i,r_y^i)$, with pseudospin $\alpha=\uparrow,\downarrow$, and the Fermi chemical potential $\mu_\alpha$. $-t_\alpha$ describes a particle hopping between the nearest neighbor sites, and $\langle i,j\rangle$ means the summation over the nearest neighbor pairs. In Eq. (\ref{eq.18}), the $s$-wave superfluid order parameter $\Delta_s({\bm r}_i)=U_s\langle c_{{\bm r}_i,\uparrow}c_{{\bm r}_i,\downarrow}\rangle$, as well as the Hartree potential $-U_sn_{-\alpha}({\bm r}_i)=-U_s\langle c_{{\bm r}_i,-\alpha}^\dagger c_{{\bm r}_i,-\alpha}\rangle$, are obtained from the mean-field approximation for the on-site Hubbard interaction $-U_s c_{{\bm r}_i,\uparrow}^\dagger c_{{\bm r}_i,\uparrow}c_{{\bm r}_i,\downarrow}^\dagger c_{{\bm r}_i,\downarrow}$, where $-U_s$ ($<0$) is the interaction strength. $V_\alpha({\bm r}_i)=V_0^\alpha({\bm r}_i/r_d)^2$ is a harmonic trap potential, where $V_0^\alpha$ is the strength of a trap potential which $\alpha$-spin atoms feel. Here, the spatial position ${\bm r}_i$ is measured from the center of the $L\times L$ square lattice, and $r_d$ is the distance between the trap center and the edge of the system. For simplicity, we take the lattice constant $a$ to be unity.
\par
In the present model, the spatial inversion symmetry is broken by the trap potential except at the trap center. For the spin rotation symmetry, it is broken when one of $t_\alpha$, $\mu_\alpha$, and $V_0^\alpha$, depends on pseudospin $\alpha=\uparrow,\downarrow$. Since all these cases satisfy Eq. (\ref{eq.15}) only when $j=z$, Eq. (\ref{eq.16}) indicates that one may only consider the possibility of the triplet pair amplitude with $S_z=0$. In the present model, this component is given by
\begin{eqnarray}
\Phi_{\rm t}^{S_z=0}({\bm r}_i,{\bm r}_j)
&=&
{1 \over \sqrt{2}}
\Bigl[
\langle c_{{\bm r}_i,\uparrow}c_{{\bm r}_j,\downarrow}\rangle
+
\langle c_{{\bm r}_i,\downarrow}c_{{\bm r}_j,\uparrow}\rangle
\Bigr].
\label{eq.25}
\end{eqnarray}
For comparison, we also consider the ordinary singlet component, given by
\begin{eqnarray}
\Phi_{\rm s}({\bm r}_i,{\bm r}_j)
&=&
{1 \over \sqrt{2}}
\Bigl[
\langle c_{{\bm r}_i,\uparrow}c_{{\bm r}_j,\downarrow}\rangle
-
\langle c_{{\bm r}_i,\downarrow}c_{{\bm r}_j,\uparrow}\rangle
\Bigr].
\label{eq.26}
\end{eqnarray}
\par
Besides the superfluid order parameter, the condensate fraction is also a fundamental quantity in the superfluid phase\cite{Yang,Salanich,Giorgini,Fukushima}, which physically describes the number of Bose-condensed Cooper pairs. In an ordinary $s$-wave superfluid state, it has the form,
\begin{equation}
N_{\rm c}^{\rm s}
=\sum_{{\bm R}=({\bm r}_i+{\bm r}_j)/2} n_{\rm c}^{\rm s}({\bm R}),
\label{eq.26b}
\end{equation}
where the local condensate fraction $n_{\rm c}^{\rm s}({\bm R})$ is directly related to the singlet pair amplitude in Eq. (\ref{eq.26}) as
\begin{equation}
n_{\rm c}^{\rm s}({\bm R})={1 \over 2N}
\sum_{{\bm r}_{\rm rel}={\bm r}_i-{\bm r}_j}
|\Phi_{\rm s}({\bm R}+{\bm r}_{\rm rel}/2,{\bm R}-{\bm r}_{\rm rel}/2)|^2.
\label{eq.26c}
\end{equation}
In addition to the singlet component of the condensate fraction $N_{\rm c}^{\rm s}$ in Eq. (\ref{eq.26b}), the present system may also have the spin-triplet component\cite{Yamaguchi}, $N_{\rm c}^{\rm t}=\sum_{{\bm R}=({\bm r}_i+{\bm r}_j)/2}n_{\rm c}^{\rm t}({\bm R})$, where
\begin{equation}
n_{\rm c}^{\rm t}({\bm R})={1 \over 2N}
\sum_{{\bm r}_{\rm rel}={\bm r}_i-{\bm r}_j}
|\Phi_{\rm t}^{S_z=0}({\bm R}+{\bm r}_{\rm rel}/2,{\bm R}-{\bm r}_{\rm rel}/2)|^2.
\label{eq.26c2}
\end{equation}
The total condensate fraction is given by $N_{\rm c}^{\rm s}+N_{\rm c}^{\rm t}$. In what follows, we simply call $N_{\rm c}^{\rm s}$ and $N_{\rm c}^{\rm t}$ the singlet and triplet condensate fraction, respectively. 
\par
\begin{figure}[t]
\begin{center}
\includegraphics[width=0.5\linewidth,keepaspectratio]{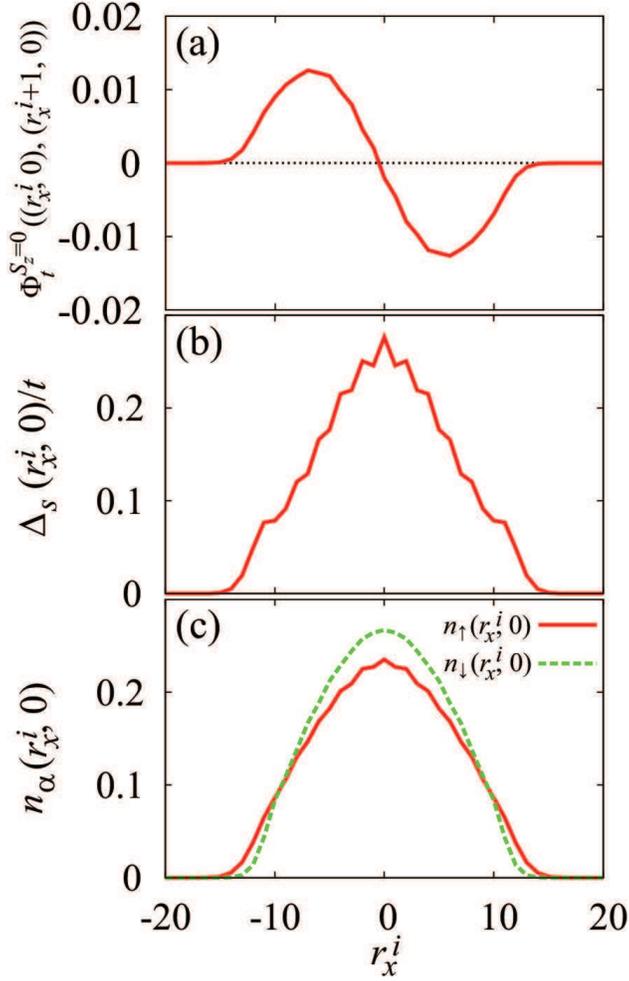}
\caption{(Color online) (a) Calculated triplet pair amplitude $\Phi_{\rm t}^{S_z=0}((r^i_x,0),(r^i_x+1,0))$ along the $x$-axis ($r^i_y=0$) in an $s$-wave superfluid Fermi gas with trap-potential difference. (b) $s$-wave superfluid order parameter $\Delta_s(r^i_x,0)$. (c) Density profile $n_\alpha(r^i_x,0)$. We take $V_0^\uparrow/V_0^\downarrow=0.5$, $t_\uparrow/t_\downarrow=1$, and $U_s/t=2.5$. This parameter set is also used in Figs. \ref{fig2}-\ref{fig4}.}
\label{fig1}
\end{center}
\end{figure}
\par
We note that the square lattice in our model does not affect the symmetry consideration in Sec. II. This is because the square lattice is invariant under the spatial inversion with respect to the center of mass position ${\bm R}=[{\bm r}_i+{\bm r}_j]/2$ of a triplet pair amplitude $\Phi_{\rm t}^{S_z}({\bm r}_i,{\bm r}_j)$. In addition, since the spin rotation symmetry is also unaffected by the crystal lattice, the necessary condition obtained in Sec. II is still valid for the present case. 
\par
As usual, we diagonalize the mean-field Hamiltonian $H_{\rm MF}$ in Eq. (\ref{eq.18}) by the Bogoliubov transformation\cite{deGennes}. Since this is a standard procedure\cite{Ohashiz}, we do not explain  the detail here, but summarize the outline in Appendix B. We numerically carry out the Bogoliubov transformation, to self-consistently determine $\Delta_s({\bm r}_i)$, $n_\alpha({\bm r}_i)$, and $\mu_\alpha$. We then evaluate the triplet pair amplitude $\Phi_{\rm t}^{S_z=0}({\bm r}_i,{\bm r}_j)$ in Eq. (\ref{eq.25}).
\par
In numerical calculations, we take the lattice size $L=41$, and $V_0^\uparrow=2t$, where $t=[t_\uparrow+t_\downarrow]/2$. To avoid lattice effects, we consider the low density region, by setting $N_\uparrow=N_\downarrow=59$ in the absence of spin imbalance (where $N_\alpha$ is the number of Fermi atoms in the $\alpha$-spin component). The total number $N=N_\uparrow+N_\downarrow$ of Fermi atoms then equals $N=118$. In this case, the particle density is at most $n_\alpha({\bm r}_i)\lesssim 0.3 \ll 1$ even in the trap center. We take a small but finite temperature $T/t=0.01$, in order to suppress effects of discrete energy levels associated with the finite system size.
\par
\begin{figure}[t]
\begin{center}
\includegraphics[width=1\linewidth,keepaspectratio]{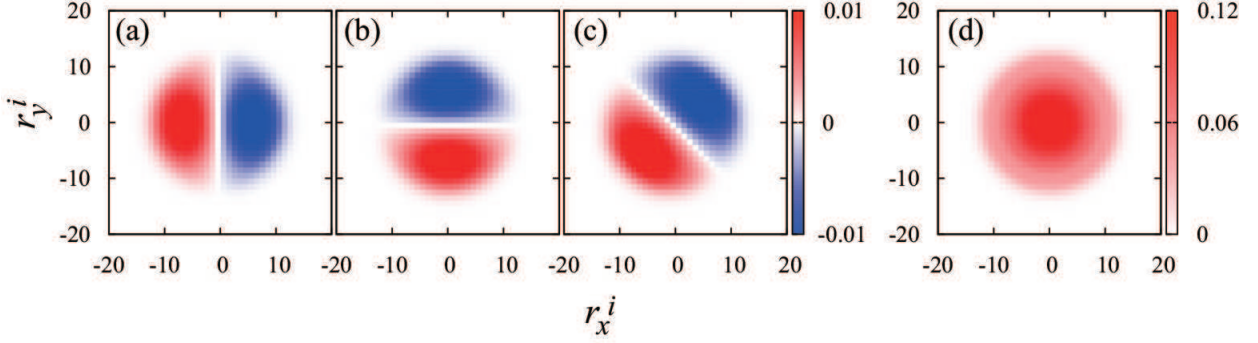}
\caption{(Color online) Calculated triplet pair amplitude $\Phi_{\rm t}^{S_z=0}({\bm r}_i,{\bm r}_j)$ in an singlet superfluid Fermi gas with trap-potential difference. (a) $\Phi_{\rm t}^{S_z=0}((r^i_x,r^i_y),(r^i_x+1,r^i_y))$. (b) $\Phi_{\rm t}^{S_z=0}((r^i_x,r^i_y),(r^i_x,r^i_y+1))$. (c) $\Phi_{\rm t}^{S_z=0}((r^i_x,r^i_y),(r^i_x+1,r^i_y+1))$. (d) Singlet pair amplitude $\Phi_{\rm s}({\bm r}_i,{\bm r}_i)$.
}
\label{fig2}
\end{center}
\end{figure}
\par
Figure \ref{fig1}(a) shows the evidence that the triplet pair amplitude with $S_z=0$ is induced in the $s$-wave superfluid state, when both the spatial inversion symmetry and the spin rotation symmetry are broken by the trap potential $V_\alpha({\bm r}_i)$. From the comparison of this figure with Figs.\ref{fig1}(b) and (c), one finds that $\Phi_{\rm t}^{S_z=0}((r_x^i+1,0),(r^i_x,0))$ appears everywhere in the gas cloud where the $s$-wave superfluid order parameter $\Delta_s(r_x^i,0)$, as well as the atom density $n_\alpha(r_x^i,0)$, are finite, except at the trap center. Since the system still has the spatial inversion symmetry at the trap center, the node structure seen in Fig. \ref{fig1}(a) agrees with the symmetry consideration in Sec. II. We emphasize that the triplet pair amplitude is not induced when $V_0^\uparrow=V_0^\downarrow$, although we do not explicitly show the result here. 
\par
Figure \ref{fig2}(a) shows that the point node seen in Fig.\ref{fig1}(a) is actually a line node along the $y$-axis. This node structure comes from the symmetry property that, while the present square-lattice model has the reflection symmetry with respect to the $y$ axis, the triplet pair amplitude $\Phi_{\rm t}^{S_z}({\bm r}_i,{\bm r}_j)$ behaves as,
\begin{equation}
\Phi_{\rm t}^{S_z}({\bm R}+{\bm r}_{\rm rel}/2,{\bm R}-{\bm r}_{\rm rel}/2)
=-
\Phi_{\rm t}^{S_z}({\bm R}-{\bm r}_{\rm rel}/2,{\bm R}+{\bm r}_{\rm rel}/2),
\label{eq.27}
\end{equation}
when ${\bm R}=[{\bm r}_i+{\bm r}_j]/2=(0,R_y)$ and ${\bm r}_{\rm rel}={\bm r}_i-{\bm r}_j=(r_{\rm rel}^x,0)$. Since the present lattice model is also invariant under the reflections with respect to the $x$ axis, as well as the lines along $y=\pm x$, the triplet pair amplitude $\Phi_{\rm t}^{S_z}({\bm r}_i,{\bm r}_j)$, with the relative vector ${\bm r}_{\rm rel}={\bm r}_i-{\bm r}_j$ being perpendicular to one of them, has the line node along the reflection line. (See Figs. \ref{fig2}(b) and (c).) In a continuum system with no lattice, the triplet pair amplitude is expected to always have the line node, which is perpendicular to the relative vector of the pair amplitude. We briefly note that such a node is not obtained in the singlet component, as shown in Fig. \ref{fig2}(d).
\par
\begin{figure}[t]
\begin{center}
\includegraphics[width=0.5\linewidth,keepaspectratio]{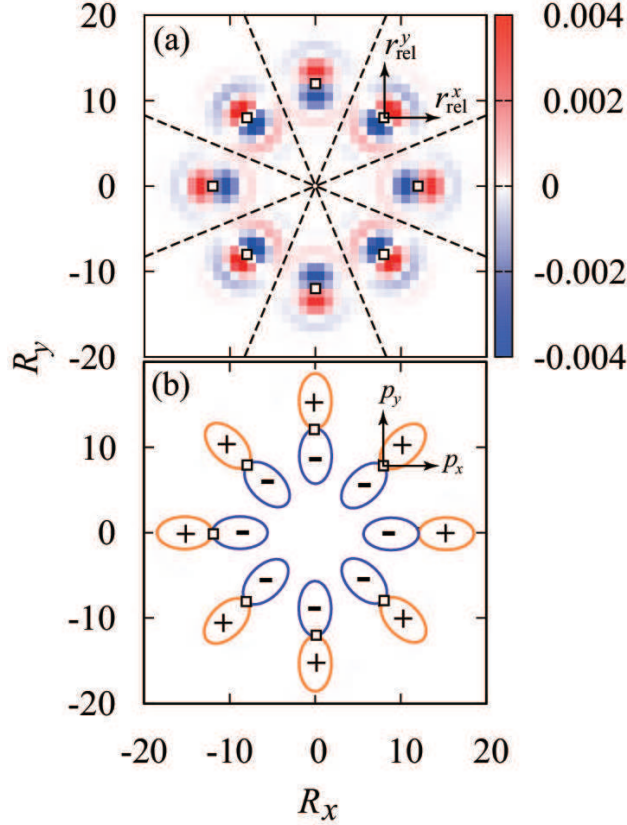}
\caption{(Color online) (a) Triplet pair amplitude $\Phi_{\rm t}^{S_z=0}({\bm r}_i,{\bm r}_j)$, as a function of the relative coordinate ${\bm r}_{\rm rel}={\bm r}_i-{\bm r}_j$. We take $V_0^\uparrow/V_0^\downarrow=0.5$. For each center of mass position ${\bm R}=[{\bm r}_i+{\bm r}_j]/2$ (open square), the pair amplitude is plotted inside the region between two dashed line, by taking ${\bm R}$ as the origin. (b) Spatial variation of synthesized $p$-wave superfluid order parameter $\Delta_p({\bm p}, {\bm R})$. In this panel, the ${\bm p}$-dependence of $\Delta_p({\bm p}, {\bm R})$ is schematically shown, being centered at ${\bm R}$ (open square).
}
\label{fig3}
\end{center}
\end{figure}
\par
Figure \ref{fig3}(a) shows the spatial structure of the triplet pair amplitude $\Phi_{\rm t}^{S_z=0}({\bm r}_i,{\bm r}_j)$ with respect to the relative coordinate ${\bm r}_{\rm rel}={\bm r}_i-{\bm r}_j$. Noting that the pairing symmetry is determined by the angular dependence in relative-momentum space, we find that the induced pair amplitude has the $p$-wave symmetry. That is, the pair amplitude has the $p_x$-wave ($p_y$-wave) symmetry, when the center of mass position is on the $x$-axis ($y$-axis). 
\par
An advantage of the cold Fermi gas system is that one can tune the pairing interaction by adjusting the threshold energy of a Feshbach resonance. Although this technique is usually used to adjust the interaction strength for a fixed interaction channel, one may also use this technique to change the interaction channel from the $s$-wave one to a $p$-wave one. In an $s$-wave superfluid Fermi gas with triplet pair amplitude shown in Fig. \ref{fig3}(a), when one suddenly change the $s$-wave interaction to the $p$-wave one\cite{OhashiP,Ho},
\begin{equation}
H_{p-\rm{wave}} = -U_p
\sum_{{\bm p},{\bm p}',{\bm q}}
{\bm p}\cdot{\bm p}'
c_{{\bm p}+{\bm q}/2,\uparrow}^\dagger
c_{-{\bm p}+{\bm q}/2,\downarrow}^\dagger
c_{-{\bm p}'+{\bm q}/2,\downarrow}
c_{{\bm p}'+{\bm q}/2,\uparrow},
\label{eq.26d}
\end{equation}
the $p$-wave superfluid order parameter,
\begin{equation}
\Delta_p({\bm p},{\bm R})=
U_p
\sum_{{\bm p}'}
{\bm p}\cdot{\bm p}'
\Phi_{\rm t}^{S_z=0}({\bm p}',{\bm R}),
\label{eq.26e}
\end{equation}
would immediately become finite. Here, $\Phi_{\rm t}^{S_z=0}({\bm p}',{\bm R})$ is the Fourier-transformed triplet pair amplitude with respect to the relative coordinate ${\bm r}_{\rm rel}$. We emphasize that this triplet pair amplitude has already existed before the change of the interaction. Thus, just after this manipulation, we expect the spatial structure of the induced $p$-wave superfluid order parameter schematically shown in Fig. \ref{fig3}(b). The $s$-wave superfluid order parameter immediately disappears because of the vanishing $s$-wave interaction ($U_s=0$), and the $s$-wave pair amplitude $\Phi_{\rm s}({\bm r}_i,{\bm r}_j)$ only remains. Thus, at least just after this manipulation, by definition, the system is in the $p$-wave superfluid state with the synthesized $p$-wave superfluid order parameter in Eq. (\ref{eq.26e}). This unconventional superfluid phase would be in the non-equilibrium state, so that we need further analyses on the time evolution of this state. However, the combined Feshbach technique with the induced triplet pair amplitude is an interesting idea to realize a $p$-wave superfluid Fermi gas.
\par
\begin{figure}[t]
\begin{center}
\includegraphics[width=0.5\linewidth,keepaspectratio]{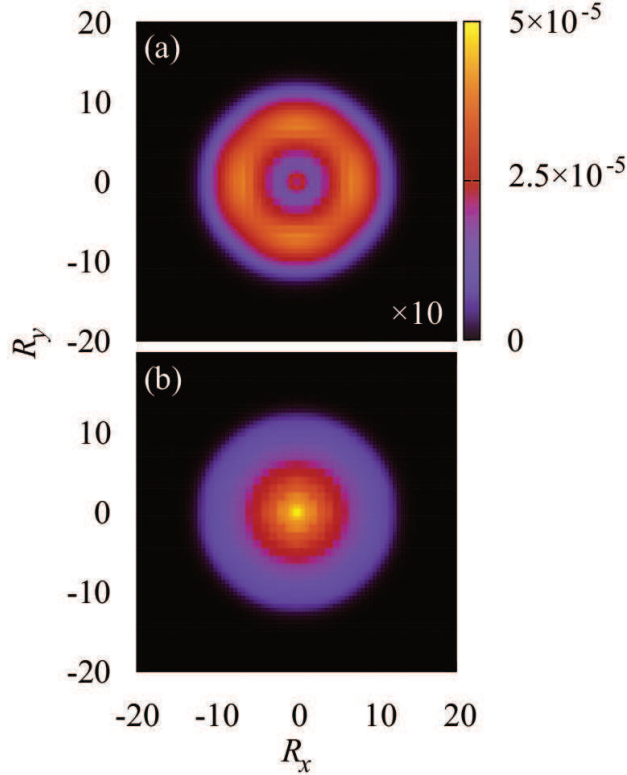}
\caption{(Color online) Local condensate fraction in an $s$-wave superfluid Fermi gas with trap-potential difference. (a) Triplet component $n_{\rm c}^{\rm t}({\bm R})$. The intensity is magnified to ten times. (b) Singlet component $n_{\rm c}^{\rm s}({\bm R})$.}
\label{fig4}
\end{center}
\end{figure}
\par
Figure \ref{fig4} shows the local condensate fraction $n_{\rm c}^{\rm s,t}({\bm R})$ in an $s$-wave superfluid Fermi gas with trap-potential imbalance. In panel (a), the triplet component $n_{\rm c}^{\rm t}({\bm R})$ is enhanced around $|{\bm R}|=6$, as well as the region near the trap center (except at ${\bm R}=0$, where the triplet condensate fraction vanishes). On the other hand, panel (b) shows that the singlet component $n_{\rm c}^{\rm s}({\bm R})$ monotonically decreases, as one goes away from the trap center. The latter behavior is consistent with the spatial variation of the $s$-wave superfluid order parameter $\Delta_s({\bm r}_i)$ shown in Fig. \ref{fig1}(b).
\par
\begin{figure}[t]
\begin{center}
\includegraphics[width=0.5\linewidth,keepaspectratio]{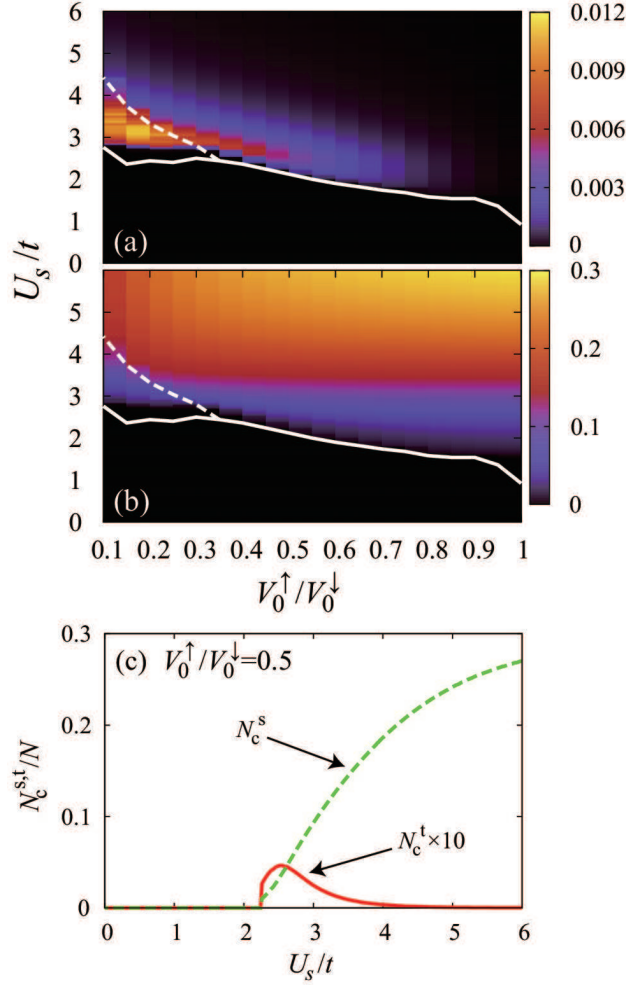}
\caption{(Color online) Condensate fraction $N_{\rm c}^{\rm s,t}=\sum_{\bm R}n_{\rm c}^{\rm s,t}({\bm R})$ in an $s$-wave superfluid Fermi gas with trap-potential difference. (a) Triplet component $N_{\rm c}^{\rm t}$. (b) Singlet component $N_{\rm c}^{\rm s}$. (c) $N_{\rm c}^{\rm t}$ and $N_{\rm c}^{\rm s}$, as functions of the interaction strength $U_s$, when $V_0^\uparrow/V_0^\downarrow=0.5$. In panels (a) and (b), the region above the solid line is in the superfluid state within our numerical accuracy. (Note that we take $T/t=0.01>0$ in our numerical calculations.) The region between the solid line and the dashed line is in the FFLO phase, being characterized by a spatially oscillating superfluid order parameter $\Delta_s({\bm r}_i)$. These lines are also drawn in Fig.\ref{fig6}.
}
\label{fig5}
\end{center}
\end{figure}
\par
The large triplet condensate fraction $n_{\rm c}^{\rm t}({\bm R})$ near the trap center seen in Fig. \ref{fig4}(a) is due to the spin imbalance ($n_\uparrow(r_x^i,0)>n_\downarrow(r^i_x,0)$) in the trap center. (See Fig. \ref{fig1}(c).) This naturally leads to the broken spin rotation symmetry through the Fermi chemical potential $\mu_\alpha$, as well as the Hartree potential $-U_sn_{-\alpha}({\bm r}_i)$ in Eq. (\ref{eq.18}). Thus, although two spin components feel almost the same trap potential ($V_\uparrow({\bm r}_i)\simeq V_\downarrow({\bm r}_i)$) around the trap center, the triplet condensate fraction is enhanced there (except at ${\bm R}=0$). 
\par
While the difference $V_\uparrow({\bm r}_i)-V_\downarrow({\bm r}_i)$ becomes remarkable as one goes away from the trap center, the spin imbalance ($n_\uparrow(r_x^i,0)-n_\downarrow(r^i_x,0)$) becomes small to eventually vanish at $|{\bm R}|\simeq 9$. (See Fig. \ref{fig1}(c).) In the outer region, the spin imbalance again occurs as $n_\uparrow(r_x^i,0)<n_\downarrow(r^i_x,0)$. These enhance $n_{\rm c}^{S_z=0}(|{\bm R}|\sim 6)$, as shown in Fig. \ref{fig4}(a).
\par
\begin{figure}[t]
\begin{center}
\includegraphics[width=0.5\linewidth,keepaspectratio]{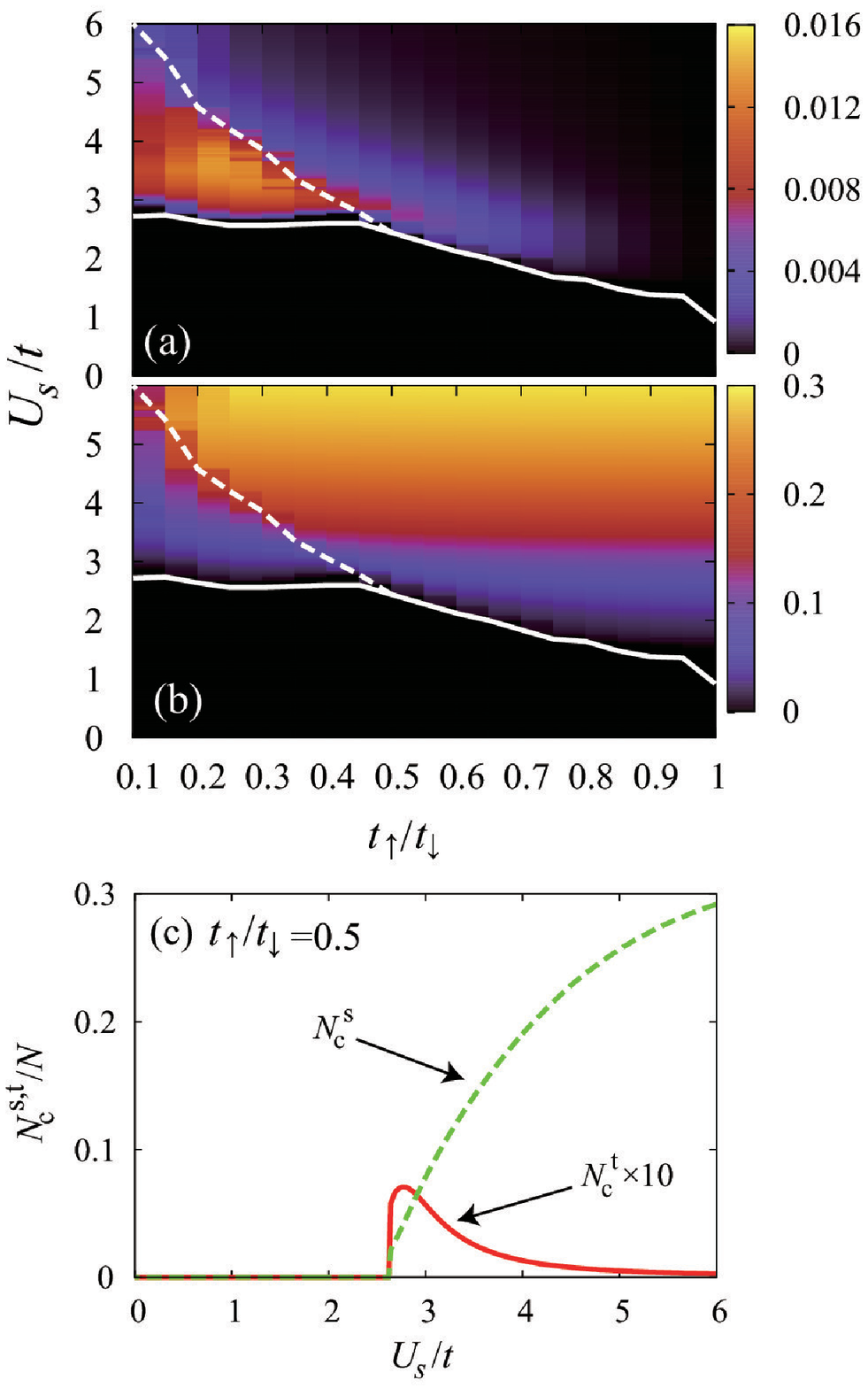}
\caption{(Color online) Condensate fraction $N_{\rm c}^{\rm s,t}$ in a trapped $s$-wave superfluid Fermi gas in the presence of mass imbalance $t_\uparrow/t_\downarrow\ne 1$. (a) Triplet component $N_{\rm c}^{\rm t}$. (b) Singlet component $N_{\rm c}^{\rm s}$. (c) $N_{\rm c}^{\rm t}$ and $N_{\rm c}^{\rm s}$, as functions of the interaction strength $U_s/t$, when $t_\uparrow/t_\downarrow=0.5$.}
\label{fig6}
\end{center}
\end{figure}
\par
Summing up the local condensate fraction $n_{\rm c}^{\rm s,t}({\bm R})$ in the gas cloud, one obtains the condensate fraction $N_{\rm c}^{\rm s,t}$ in Fig. \ref{fig5}. As expected, panel (a) shows that the triplet component $N_{\rm c}^{\rm t}$ is enhanced when $V_0^\uparrow/V_0^\downarrow\ll 1$. We also find that $N_{\rm c}^{\rm t}$ becomes large in the intermediate coupling regime but becomes small when $U_s/t\gg 1$. In the strong coupling regime, most Fermi atoms form singlet molecules, which suppresses effects of broken inversion and spin rotation symmetry. Indeed, Fig. \ref{fig5}(b) shows that the singlet component $N_{\rm c}^{\rm s}$ monotonically increases with increasing the interaction strength $U_s$. To clearly see the difference between $N_{\rm c}^{\rm t}$ and $N_{\rm c}^{\rm s}$, Fig. \ref{fig5}(c) shows these quantities as functions of the interaction strength $U_s$.
\par
In Figs. \ref{fig5}(a) and (b), one sees the FFLO (Fulde-Ferrell-Larkin-Ovchinnikov) phase\cite{FFLO1,FFLO2,FFLO3,Mizushima}. In this regard, since we are dealing with a two-dimensional lattice model within the simple mean-field theory, it is unclear whether or not the FFLO phase still remains in a realistic three-dimensional continuum Fermi gas\cite{OhashiJPSJ}. However, since Fig. \ref{fig5}(a) indicates that the triplet condensate fraction is also induced in the ordinary BCS region, we find that the FFLO state is not necessary for the triplet pair amplitude to appear. 
\par
\begin{figure}[t]
\begin{center}
\includegraphics[width=0.5\linewidth,keepaspectratio]{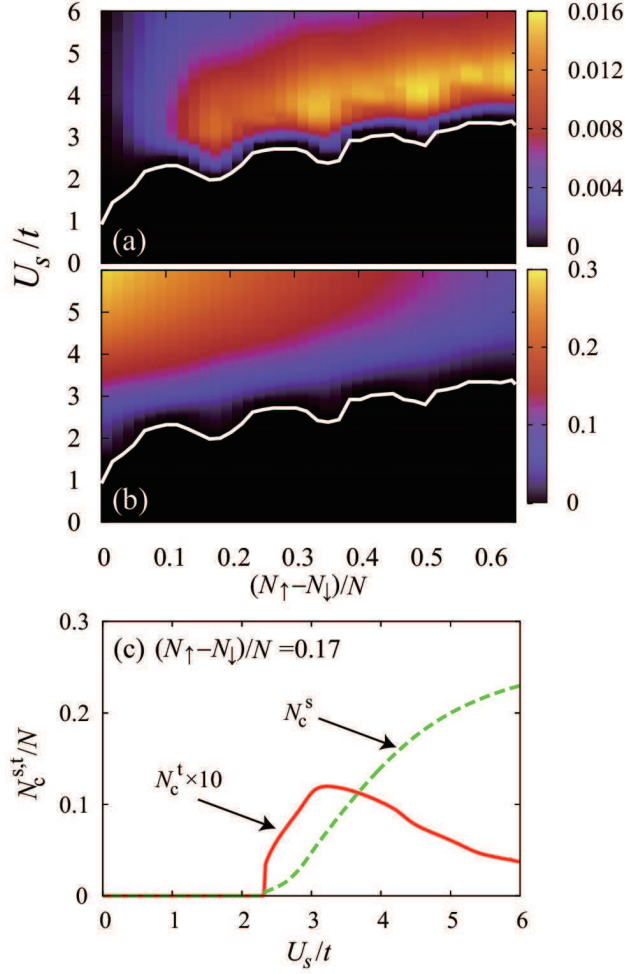}
\caption{(Color online) Condensate fraction $N_{\rm c}^{\rm s,t}$ in a trapped $s$-wave superfluid Fermi gas with spin imbalance $N_\uparrow/N_\downarrow\ne 1$. (a) Triplet component $N_{\rm c}^{\rm t}$. (b) Singlet component $N_{\rm c}^{\rm s}$. (c) $N_{\rm c}^{\rm t}$ and $N_{\rm c}^{\rm s}$, as functions of the interaction strength $U_s/t$, when $[N_\uparrow-N_\downarrow]/N=0.17$ ($N_\uparrow-N_\downarrow=20$). In panels (a) and (b), the region above the solid line is in the superfluid phase. In the superfluid region shown in this figure, the superfluid order parameter in the outer region of the gas cloud always exhibits a FFLO-type oscillation in the radial direction.}
\label{fig7}
\end{center}
\end{figure}

\begin{figure}[t]
\begin{center}
\includegraphics[width=0.5\linewidth,keepaspectratio]{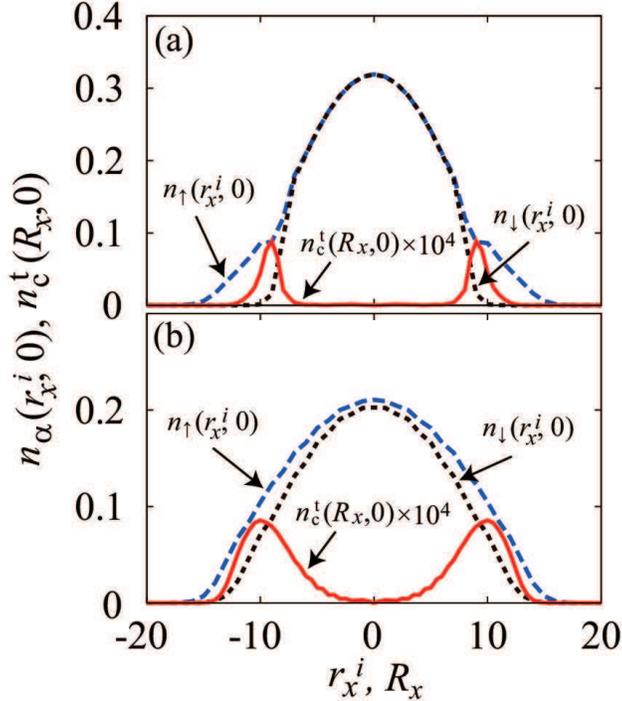}
\caption{(Color online) Calculated density profile $n_\alpha(r_x^i,0)$ and the triplet condensate fraction $n_{\rm c}^{\rm t}(R_x,0)$ in a trapped $s$-wave superfluid Fermi gas with spin imbalance. (a) $U_s/t=6$. (b) $U_s/t=3$. We set $(N_\uparrow-N_\downarrow)/N=0.17$.}
\label{fig8}
\end{center}
\end{figure}
\par
Figure \ref{fig6} confirms that the triplet pair amplitude $\Phi_{\rm t}^{S_z=0}({\bm r}_i,{\bm r}_j)$ is also induced, when the spin rotation symmetry is broken by mass imbalance ($t_\uparrow/t_\downarrow\ne 1$). In addition, Fig. \ref{fig7} shows that this phenomenon also occurs in a trapped $s$-wave superfluid Fermi gas with spin imbalance ($N_\uparrow/N_\downarrow\ne 1$)\cite{note}. In the latter case, the Fermi chemical potential $\mu_\alpha$ depends on pseudospin $\alpha=\uparrow,\downarrow$, which breaks the spin rotation symmetry\cite{note2}. 
\par
In the presence of spin imbalance, the phase separation is known to occur\cite{Ketterle,Hulet,Shin}, where the superfluid region in the trap center is surrounded by excess atoms. Figure \ref{fig8}(a) shows this case. In this panel, since the spin imbalance is almost absent around the trap center, the triplet condensate fraction is suppressed there, compared to the case when the phase separation does not occur (panel (b)). In addition, the region around the edge of the gas cloud is highly spin polarized, so that the triplet pair amplitude is also suppressed there. As a result, when the phase separation occurs, the triplet pair amplitude is localized around the edge of the gas cloud of the minority component ($\alpha=\downarrow$), as shown in Fig. \ref{fig8}(b).
\par
\section{Summary}
To summarize, we have discussed the possibility of inducing a triplet pair amplitude in a trapped $s$-wave superfluid Fermi gas. Using symmetry considerations, we clarified that both the broken spatial inversion symmetry and a broken spin rotation symmetry are necessary for this phenomenon to occur. We numerically confirmed that a triplet pair amplitude is induced when this condition is satisfied, within the mean-field theory for a two-dimensional lattice model. In this confirmation, we considered the three cases with (1) trap-potential difference, (2) mass imbalance, and (3) spin imbalance. In the first case, we showed that the induced triplet pair amplitude is dominated by a $p$-wave symmetry. Among the above three cases, a trapped $s$-wave superfluid Fermi gas with spin imbalance has been realized\cite{Ketterle,Hulet,Shin}. Thus, our results imply that a triplet pair amplitude is already present in this system, although there is no experimental evidence yet.
\par
Since the symmetry of a Fermi superfluid is fully determined by the symmetry of the superfluid order parameter, the induction of a triplet pair amplitude does not immediately mean the realization of a triplet Fermi superfluid. In the present case, the system is still in the $s$-wave superfluid state which is characterized by the $s$-wave superfluid order parameter, even in the presence of a triplet pair amplitude. Under this situation, however, when one suddenly changes the $s$-wave pairing interaction to an appropriate $p$-wave one, the product of the $p$-wave interaction and the triplet pair amplitude that has been induced before this manipulation may immediately give a finite $p$-wave superfluid order parameter. Since the $s$-wave superfluid order parameter vanishes, by definition, we have a $p$-wave superfluid state, characterized by this $p$-wave superfluid parameter. The change of the interaction would be possible by using the Feshbach resonance technique.
\par
In cold Fermi gas physics, although the realization of a $p$-wave superfluid state is a crucial challenge, current experiments are facing various difficulties originating from a $p$-wave interaction, such as three particle loss\cite{3loss1,3loss2,3loss3}, as well as the dipolar relaxation\cite{Gaebler1}. In this regard, the above idea may avoid these difficulties to some extent, because the triplet pair amplitude is prepared in an $s$-wave superfluid Fermi gas with no $p$-wave interaction. In addition, since we can start from a finite value of the $p$-wave superfluid order parameter, the system would be in the $p$-wave superfluid state for a while, until it is strongly damaged by the particle loss and dipolar relaxation after the $p$-wave interaction is introduced. In this sense, the induction of a triplet pair amplitude discussed in this paper is important, not only as a fundamental physical phenomenon, but also on the viewpoint of the challenge toward the realization of a $p$-wave superfluid Fermi gas.
\par
In this paper, we have treated a lattice model to simply confirm the induction of a triplet pair amplitude. To quantitatively evaluate this quantity, we need to extend the present analyses to a realistic continuum Fermi superfluid. To assess the idea that one produces a $p$-wave superfluid Fermi gas from the induced triplet pair amplitude, it is also important to clarify the time evolution of the $p$-wave superfluid order parameter after the $s$-wave interaction is replaced by a $p$-wave one. These problems remain as our future problems. Since the pair amplitude always exists in a Fermi superfluid, our results would be useful for the study of this fundamental quantity in cold Fermi gas physics.
\par
\begin{acknowledgements}
YO thanks N. Yamamoto for useful comments. This work was supported by KiPAS project in Keio University. YE and RH were supported by a Grant-in-Aid for JSPS fellows. YO was supported by Grant-in-Aid for Scientific research from MEXT in Japan (25105511, 25400418, 15H00840). 
\end{acknowledgements}
\par
\appendix
\section{Triplet pair amplitude in a spin-orbit coupled uniform $s$-wave superfluid Fermi gas}
\par
We consider a uniform $s$-wave superfluid Fermi gas with a spin-orbit interaction. The model Hamiltonian is given by
\begin{equation}
H=\sum_{{\bm p},\alpha}\xi_{\bm p}
c^\dagger_{{\bm p},\alpha}c_{{\bm p},\alpha}+H_{\rm so}
-U_s\sum_{{\bm p},{\bm p}',{\bm q}}
c^\dagger_{{\bm p}+{\bm q}/2,\uparrow}
c^\dagger_{-{\bm p}+{\bm q}/2,\downarrow}
c_{-{\bm p}'+{\bm q}/2,\downarrow}
c_{{\bm p}'+{\bm q}/2,\uparrow}.
\label{eq.a1}
\end{equation}
Here, $c^\dagger_{{\bm p},\alpha}$ is a creation operator of a Fermi atom with the kinetic energy $\xi_{\bm p}=p^2/(2m)-\mu$, measured from the chemical potential $\mu$. The antisymmetric spin-orbit interaction $H_{\rm so}$ has the form\cite{Yamaguchi},
\begin{equation}
H_{\rm so}=
\sum_{{\bm p},\alpha,\alpha'}
c^\dagger_{{\bm p},\alpha}
h_{\rm so}^{\alpha,\alpha}({\bm p})
c_{{\bm p},\alpha'},
\label{eq.a2}
\end{equation}
where ${\hat h}_{\rm so}({\bm p})=\{h_{\rm so}^{\alpha,\alpha'}\}$ is assumed as
\begin{equation}
{\hat h}_{\rm so}({\bm p})=
\lambda_\perp[p_x\sigma_x+p_y\sigma_y]+\lambda_z\sigma_z.
\label{eq.a3}
\end{equation}
Here, $\lambda_\perp$ and $\lambda_z$ are spin-orbit couplings. 
\par
We assume that the system is in the $s$-wave superfluid state with the superfluid order parameter $\Delta_s=U_s\sum_{\bm p}\langle c_{{\bm p}\uparrow}c_{-{\bm p}\downarrow}\rangle$, We also assume that any other spontaneous symmetry breaking is absent.
\par
In momentum space, the spatial inversion ${\hat P}$ is described as ${\tilde c}_{{\bm p},\alpha}={\hat P}c_{{\bm p},\alpha}{\hat P}^{-1}=c_{-{\bm p},\alpha}$. Under this operation, each term in Eq. (\ref{eq.a1}) is invariant, except for the spin-orbit interaction, which is transformed as
\begin{equation}
{\tilde H}_{\rm so}=
{\hat P}H_{\rm so}{\hat P}^{-1}
=
\sum_{{\bm p},\alpha,\alpha'}
c^\dagger_{-{\bm p},\alpha}
h_{\rm so}^{\alpha,\alpha'}({\bm p})
c_{-{\bm p},\alpha'}
=
\sum_{{\bm p},\alpha,\alpha'}
c^\dagger_{{\bm p},\alpha}
h_{\rm so}^{\alpha,\alpha'}(-{\bm p})
c_{{\bm p},\alpha'}
=
-H_{\rm so}.
\label{eq.a4}
\end{equation}
Thus, the spin-orbit interaction $H_{\rm so}$ in Eq. (\ref{eq.a2}) breaks the inversion symmetry.
\par
For the spin rotation ${\hat R}({\bm \theta})$, the three $\pi$ rotations (${\bm \theta}={\bm \theta}_\pi^j$, $j=x,y,z$) corresponding to Eq. (\ref{eq.13}) are given by
\begin{eqnarray}
\left(
\begin{array}{c}
{\tilde c}_{{\bm p}\uparrow}\\
{\tilde c}_{{\bm p}\downarrow}
\end{array}
\right)_{{\bm \theta}={\bm \theta}_\pi^j}
=
i\sigma_j
\left(
\begin{array}{c}
c_{{\bm p}\uparrow}\\
c_{{\bm p}\downarrow}
\end{array}
\right)~~~(j=x,y,z).
\label{eq.a5}
\end{eqnarray}
When $\lambda_\perp=0$ and $\lambda_z\ne 0$ (single component spin orbit interaction), Eq. (\ref{eq.a1}) is not invariant under the spin rotations ${\hat R}({\bm \theta}_\pi^x)$ and ${\hat R}({\bm \theta}_\pi^y)$, because the spin-orbit interaction $H_{\rm so}$ is transformed as
\begin{equation}
{\tilde H}_{\rm so}
=
{\hat R}({\bm \theta}_\pi^{x,y})H_{\rm so}
{\hat R}^{-1}({\bm \theta}_\pi^{x,y})
=\lambda_z
\sum_{{\bm p},\alpha,\alpha'}
{\tilde c}^\dagger_{{\bm p},\alpha}
\sigma_z^{\alpha,\alpha'}
{\tilde c}_{{\bm p},\alpha'}
=-\lambda_z
\sum_{{\bm p},\alpha,\alpha'}
c^\dagger_{{\bm p},\alpha}
\sigma_z^{\alpha,\alpha'}
c_{{\bm p},\alpha'}
=
-H_{\rm so}.
\label{eq.a6}
\end{equation}
Thus, Eq. (\ref{eq.a1}) is invariant only under the $\pi$ rotation with ${\bm \theta}={\bm \theta}_\pi^z$. Noting that the triplet pair amplitude
\begin{eqnarray}
\Phi_{\rm t}^{S_z}({\bm p})=
\left\{
\begin{array}{ll}
\langle c_{{\bm p}\uparrow}c_{-{\bm p}\uparrow}\rangle &~~~(S_z=1),\\
{1 \over \sqrt{2}}
\left[
\langle c_{{\bm p}\uparrow}c_{-{\bm p}\downarrow}\rangle
+
\langle c_{{\bm p}\downarrow}c_{-{\bm p}\uparrow}\rangle
\right]
 &~~~(S_z=0),\\
\langle c_{{\bm p}\downarrow}c_{-{\bm p}\downarrow}\rangle &~~~(S_z=-1),\\
\end{array}
\right.
\label{eq.a7}
\end{eqnarray}
is transformed under the three $\pi$ rotations as
\begin{eqnarray}
\left(
\begin{array}{c}
{\tilde \Phi}_{\rm t}^{S_z=1}({\bm p}) \\
{\tilde \Phi}_{\rm t}^{S_z=0}({\bm p}) \\
{\tilde \Phi}_{\rm t}^{S_z=-1}({\bm p})
\end{array}
\right)
=
\left\{
\begin{array}{ll}
\left(
\begin{array}{c}
-\Phi_{\rm t}^{S_z=-1}({\bm p}) \\
-\Phi_{\rm t}^{S_z=0}({\bm p}) \\
-\Phi_{\rm t}^{S_z=1}({\bm p})
\end{array}
\right)& ({\bm \theta}={\bm \theta}_\pi^x),\\
\left(
\begin{array}{c}
\Phi_{\rm t}^{S_z=-1}({\bm p}) \\
-\Phi_{\rm t}^{S_z=0}({\bm p}) \\
\Phi_{\rm t}^{S_z=1}({\bm p})
\end{array}
\right)& ({\bm \theta}={\bm \theta}_\pi^y,)\\
\left(
\begin{array}{c}
-\Phi_{\rm t}^{S_z=1}({\bm p}) \\
\Phi_{\rm t}^{S_z=0}({\bm p}) \\
-\Phi_{\rm t}^{S_z=-1}({\bm p})
\end{array}
\right)& ({\bm \theta}={\bm \theta}_\pi^z),\\
\end{array}
\right.
\label{eq.a8}
\end{eqnarray}
we find that only $\Phi_{\rm t}^{S_z=0}({\bm p})$ may be induced. Indeed, Ref.\cite{Yamaguchi} shows that it is induced in this case.
\par
When $\lambda_\perp\ne 0$, the spin-orbit interaction $H_{\rm so}$ is not invariant under any $\pi$ rotations ${\hat R}({\bm \theta}_\pi^{x,y,z})$. Within this analysis, one concludes that all the triplet pair amplitudes in Eq. (\ref{eq.a7}) may be induced. However, within the mean-field theory, Ref.\cite{Yamaguchi} shows that the component with $S_z=0$ is not induced when $\lambda_\perp\ne 0$ and $\lambda_z=0$. This is because, in this two-component case, the mean-field BCS Hamiltonian,
\begin{equation}
H_{\rm BCS}=\sum_{{\bm p},\alpha}\xi_{\bm p}
c^\dagger_{{\bm p},\alpha}c_{{\bm p},\alpha}+H_{\rm so}
+\Delta_s\sum_{\bm p}
\Bigl[
c^\dagger_{{\bm p},\uparrow}c^\dagger_{-{\bm p},\downarrow}+{\rm h.c}
\Bigr],
\label{eq.a9}
\end{equation}
is invariant under the momentum dependent $\pi$ spin-rotation which is followed by the $U(1)$ gauge transformation, given by
\begin{eqnarray}
\left(
\begin{array}{c}
{\tilde c}_{{\bm p}\uparrow}\\
{\tilde c}_{{\bm p}\downarrow}
\end{array}
\right)
=
e^{-i{\pi \over 2}}\times e^{i{\pi \over 2}{\hat {\bm p}_\perp}\cdot{\bm \sigma}}
\left(
\begin{array}{c}
c_{{\bm p}\uparrow}\\
c_{{\bm p}\downarrow}
\end{array}
\right),
\label{eq.a10}
\end{eqnarray}
where ${\hat {\bm p}_\perp}=(p_x,p_y)/\sqrt{p_x^2+p_y^2}$. In this case, the triplet pair amplitude with $S_z=0$ is transformed as $\Phi_{\rm t}^{S_z=0}({\bm p})\to -\Phi_{\rm t}^{S_z=0}({\bm p})$, so that one finds $\Phi_{\rm t}^{S_z=0}({\bm p})=0$, as obtained in Ref.\cite{Yamaguchi}.
\par
\par
\section{Diagonalization of the BCS Hamiltonian in Eq. (\ref{eq.18})}
\par
The mean-field BCS Hamiltonian in Eq. (\ref{eq.18}) can be diagonalized by the Bogoliubov transformation in real space, given by
\begin{eqnarray}
\left(
\begin{array}{c}
c_{{\bm r}_1,\uparrow} \\
\vdots \\
c_{{\bm r}_{L^2},\uparrow} \\
c^\dagger_{{\bm r}_1,\downarrow} \\
\vdots \\
c^\dagger_{{\bm r}_{L^2},\downarrow}
\end{array}
\right)
=
{\hat W}
\left(
\begin{array}{c}
\gamma_{1} \\
\vdots \\
\gamma_{L^2} \\
\gamma_{L^2+1} \\
\vdots \\
\gamma_{2L^2}
\end{array}
\right).
\label{eq.19}
\end{eqnarray}
Here, $\hat{W}$ is a $2L^2\times 2L^2$ orthogonal matrix, which is chosen so that $H_{\rm MF}$ in Eq. (\ref{eq.18}) can be diagonalized as 
\begin{eqnarray}
H_{\rm MF}=\sum_{j=1}^{2L^2}E_j\gamma_j^\dagger\gamma_j,
\label{eq.20}
\end {eqnarray}
where $E_j$ is a Bogoliubov single-particle excitation energy. After the diagonalization, the superfluid order parameter $\Delta_s({\bm r}_i)$, as well as the number density $n_\alpha({\bm r}_i)=\langle c_{{\bm r}_i,\alpha}^\dagger c_{{\bm r}_i,\alpha}\rangle$, are evaluated as, respectively
\begin{eqnarray}
\Delta_s({\bm r}_i)=U_s\sum_{j=1}^{2L^2}
W_{i,j}W_{i+L^2,j}f(-E_j)
\label{eq.21}
\end{eqnarray}
\begin{eqnarray}
n_\uparrow({\bm r}_i)=\sum_{j=1}^{2L^2}W_{i,j}^2 f(E_j),
\label{eq.22}
\end{eqnarray}
\begin{eqnarray}
n_\downarrow({\bm r}_i)=\sum_{j=1}^{2L^2}W_{i+L^2,j}^2f(-E_j),
\label{eq.23}
\end{eqnarray}
where $f(E)=1/[e^{\beta E_j}+1]$ is the Fermi distribution function. The number $N_\alpha$ of Fermi atoms in the $\alpha$-spin component is given by
\begin{equation}
N_\alpha=\sum_{i=1}^{L^2}n_\alpha({\bm r}_i).
\label{eq.24}
\end{equation}
We numerically calculate Eqs. (\ref{eq.19}), and (\ref{eq.21})-(\ref{eq.24}), to self-consistently determine $\Delta_s({\bm r}_i)$, $n_\alpha({\bm r}_i)$, and $\mu_\alpha$. The triplet pair amplitude $\Phi_{\rm t}^{S_z=0}({\bm r}_i,{\bm r}_j)$ in Eq. (\ref{eq.25}), as well as the singlet pair amplitude $\Phi_{\rm s}({\bm r}_i,{\bm r}_j)$ in Eq. (\ref{eq.26}), are then calculated as, respectively,
\begin{eqnarray}
\Phi_{\rm t}^{S_z=0}({\bm r}_i,{\bm r}_j)
=\sum_{k=1}^{2L^2}
\Bigl[
W_{i,k}W_{j+L^2,k}f(-E_k)+W_{i+L^2,k}W_{j,k}f(E_k)
\Bigr],
\label{eq.25a}
\end{eqnarray}
\begin{eqnarray}
\Phi_{\rm s}({\bm r}_i,{\bm r}_j)
=\sum_{k=1}^{2L^2}
\Bigl[
W_{i,k}W_{j+L^2,k}f(-E_k)-W_{i+L^2,k}W_{j,k}f(E_k)
\Bigr].
\label{eq.26a}
\end{eqnarray}
\par



\begin{thebibliography}{99}
\bibitem{Jin} C. A. Regal, M. Greiner, and D. S. Jin, Phys. Rev. Lett. \textbf{92}, 040403 (2004).
\bibitem{Zwierlein} M. W. Zwierlein, C. A. Stan, C. H. Schunck, S. M. F. Raupach, A. J. Kerman, and W. Ketterle, Phys. Rev. Lett. {\bf 92}, 120403 (2004).
\bibitem{Kinast} J. Kinast, S. L. Hemmer, M. E. Gehm, A. Turlapov, and J. E. Thomas, Phys. Rev. Lett. \textbf{92}, 150402 (2004).
\bibitem{Bartenstein} M. Bartenstein, A. Altmeyer, S. Riedl, S. Jochim, C. Chin, J. H. Denschlag, and R. Grimm, Phys. Rev. Lett. {\bf 92}, 203201 (2004).
\bibitem{Ketterle} M. W. Zwierlein, A. Schirotzek, C. H. Schunck, and W. Ketterle, Science, {\bf 311}, 492 (2006).
\bibitem{Hulet} G. B. Partridge, W. Li, R. I. Kamar, Y.-A. Liao, and R. G. Hulet, Science {\bf 311}, 503 (2006).
\bibitem{Shin} Y. Shin, C. H. Schunck, A. Schirotzek, and W. Ketterle, Nature (London) \textbf{451}, 689 (2008).
\bibitem{Chin} For a review, see, C. Chin, R. Grimm, P. Julienne, and E. Tiesinga, Rev. Mod. Phys. {\bf 82}, 1225 (2010).
\bibitem{Regal} C. A. Regal, C. Ticknor, J. L. Bohn, and D. S. Jin, Phys. Rev. Lett. {\bf 90}, 053201 (2003).
\bibitem{Ticknor} C. Ticknor, C. A. Regal, D. S. Jin, and J. L. Bohn, Phys. Rev. A {\bf 69}, 042712 (2004).
\bibitem{Zhang} J. Zhang, E. G. M. van Kempen, T. Bourdel, L. Khaykovich, J. Cubizolles, F. Chevy, M. Teichmann, L. Tarruell, S. J. J. M. F. Kokkelmans, and C. Salomon, Phys. Rev. A {\bf 70}, 030702(R) (2004).
\bibitem{Nakasuji} T. Nakasuji, J. Yoshida, and T. Mukaiyama, Phys. Rev. A {\bf 88}, 012710 (2013).
\bibitem{Gaebler2007} J. P. Gaebler, J. T. Stewart, J. L. Bohn, and D. S. Jin, Phys. Rev. Lett. {\bf 98}, 200403 (2007).
\bibitem{Inada2008} Y. Inada, M. Horikoshi, S. Nakajima, M. Kuwata-Gonokami, M. Ueda, and T. Mukaiyama, Phys. Rev. Lett. {\bf 101}, 100401 (2008).
\bibitem{Fuchs2008} J. Fuchs, C. Ticknor, P. Dyke, G. Veeravalli, E. Kuhnle, W. Rowlands, P. Hannaford, and C. J. Vale, Phys. Rev. A {\bf 77}, 053616 (2008)
\bibitem{Yamaguchi} T. Yamaguchi and Y.Ohashi, arXiv:1504.03835.
\bibitem{SOC0} For a review, see, J. Dalibard, F. Gerbier, G. Juzeli${\bar {\rm u}}$nas, and P. \"Ohberg, Rev. Mod. Phys. \textbf{83}, 1523 (2011). 
\bibitem{Rb-SOC1} Y.-J. Lin, R. L. Compton, A. R. Perry, W. D. Phillips, J. V. Porto, and I. B. Spielman, Phys. Rev. Lett. \textbf{102}, 130401 (2009). 
\bibitem{Rb-SOC2} Y.-J. Lin, R. L. Compton, K. Jim\'{e}nez-Garc\'{i}a, J. V. Porto, and I. B. Spielman, Nature \textbf{462}, 628 (2009). 
\bibitem{Rb-SOC3} Y.-J. Lin, R. L. Compton, K. Jim\'{e}nez-Garc\'{i}a, W. D. Phillips, J. V. Porto, and I. B. Spielman, Nat. Phys. \textbf{7}, 531 (2011). 
\bibitem{Rb-SOC4} Y.-J. Lin, K. Jim\'{e}nez-Garc\'{i}a, and I. B. Spielman, Nature \textbf{471}, 83 (2011). 
\bibitem{Li-SOC} L. W. Cheuk, A. T. Sommer, Z. Hadzibabic, T. Yefsah, W. S. Bakr, and M. W. Zwierlein, Phys. Rev. Lett. \textbf{109}, 095302 (2012).
\bibitem{K-SOC} P. Wang, Z.-Q. Yu, Z. Fu, J. Miao, L. Huang, S. Chai, H. Zhai, and J. Zhang, Phys. Rev. Lett. \textbf{109}, 095301 (2012).
\bibitem{coldSOC5} J. D. Sau, R. Sensarma, S. Powell, I. B. Spielman, S. DasSarma, Phys. Rev. B \textbf{83}, 140510(R) (2011).  
\bibitem{coldSOC6} B. M. Anderson, I. B. Spielman, and G. Juzeli\={u}nas, Phys. Rev. Lett. \textbf{111}, 125301 (2013). 
\bibitem{Bauer2004} E. Bauer, G. Hilscher, H. Michor, Ch. Paul, E. W. Scheidt, A. Gribanov, Yu. Seropegin, H. No{\"e}l, M. Sigrist, and P. Rogl, Phys. Rev. Lett. \textbf{92}, 027003 (2004).
\bibitem{Sigrist} V. P. Mineev and M. Sigrist, in {\it Non-centrosymmetric Superconductors}, edited by E. Bauer and M. Sigrist (Springer-Verlag, Berlin, 2012) Chapter 4.
\bibitem{Yuan} H. Q. Yuan, D. F. Agterberg, N. Hayashi, P. Badica, D. Vandervelde, K. Togano, M. Sigrist, and M. B. Salamon, Phys. Rev. Lett. \textbf{97}, 017006 (2006).
\bibitem{Nascimbene} S. Nascimb\`ene, N. Navon, S. Pilati, F. Chevy, S. Giorgini, A. Georges, and C. Salomon, Phys. Rev. Lett. \textbf{106}, 215303 (2011). 
\bibitem{Kashimura} T. Kashimura, R. Watanabe, and Y. Ohashi, Phys. Rev. A \textbf{86}, 043622 (2012), {\it ibid}, \textbf{89}, 013618 (2014). 
\bibitem{Taglieber2008} M. Taglieber, A.-C. Voigt, T. Aoki, T. W. H{\"a}nsch, and K. Dieckmann, Phys. Rev. Lett. {\bf 100}, 010401 (2008).
\bibitem{Wille2008} E. Wille, F. M. Spiegelhalder, G. Kerner, D. Naik, A. Trenkwalder, G. Hendl, F. Schreck, R. Grimm, T. G. Tiecke, J. T. M. Walraven, S. J. J. M. F. Kokkelmans, E. Tiesinga, and P. S. Julienne, Phys. Rev. Lett. {\bf 100}, 053201 (2008).
\bibitem{Spiegelhalder2009} F. M. Spiegelhalder, A. Trenkwalder, D. Naik, G. Hendl, F. Schreck, and R. Grimm, Phys. Rev. Lett. {\bf 103}, 223203 (2009).
\bibitem{Takemori2012} N. Takemori and A. Koga J. Phys. Soc. Jpn. {\bf 81}, 063002 (2012).
\bibitem{Lin2006} G.-D. Lin, W. Yi and L. -M. Duan, Phys. Rev. A {\bf 74}, 031604(R) (2006).
\bibitem{Dao2007} T.-L. Dao, A. Georges, and M. Capone, Phys. Rev. B {\bf 76}, 104517 (2007).
\bibitem{Cazalilla2005} M. A. Cazalilla, A. F. Ho, and T. Giamarchi, Phys. Rev. Lett. {\bf 95}, 226402 (2005).
\bibitem{Hanai2013} R. Hanai, T. Kashimura, R. Watanabe, D. Inotani, and Y. Ohashi, Phys. Rev. A {\bf 88}, 053621 (2013).
\bibitem{Hanai2014} R. Hanai and Y. Ohashi, Phys. Rev. A {\bf 90}, 043622 (2014).
\bibitem{Hu2011} L. Jiang, X. J. Liu, H. Hu, and H. Pu, Phys. Rev. A {\bf 84}, 063618 (2011).
\bibitem{2B1} H. Hu, L. Jiang, X.-J. Liu, and H. Pu, Phys. Rev. Lett. {\bf 107}, 195304 (2011). 
\bibitem{Yang} C. N. Yang, Rev. Mod. Phys. \textbf{34}, 694 (1962).
\bibitem{Salanich} L. Salasnich, N. Manini, and A. Parola, Phys. Rev. A \textbf{72}, 023621 (2005).
\bibitem{Giorgini} G. E. Astrakharchik, J. Boronat, J. Casulleras, and S. Giorgini, Phys. Rev. Lett. \textbf{95}, 230405 (2005).
\bibitem{Fukushima} N. Fukushima, Y. Ohashi, E. Taylor, and A. Griffin, Phys. Rev. A \textbf{75}, 033609 (2007).
\bibitem{deGennes} P. G. de Gennes, {\it Superconductivity of Metals and Alloys} (Addison-Wesley, New York, 1989).
\bibitem{Ohashiz} Y. Ohashi, Phys. Rev. A {\bf 83}, 063611 (2011).
\bibitem{OhashiP} Y. Ohashi, Phys. Rev. Lett. {\bf 94}, 050403 (2005).
\bibitem{Ho} T.-L. Ho and R. B. Diener, Phys. Rev. Lett. {\bf 94}, 090402 (2005).
\bibitem{FFLO1} P. Fulde, and R. A. Ferrell, Phys. Rev. {\bf 135}, A550 (1964).
\bibitem{FFLO2} A. I. Larkin, and Yu. N. Ovchinnikov, Sov. Phys. JETP {\bf 20}, 762 (1965).
\bibitem{FFLO3} S. Takada, and T. Izuyama, Prog. Theor. Phys. {\bf 41}, 635 (1969).
\bibitem{Mizushima} T. Mizushima, M. Ichioka, and K. Machida, J. Phys. Soc. Jpn. \textbf{76}, 104006 (2007).
\bibitem{OhashiJPSJ} Y. Ohashi, J. Phys. Soc. Jpn. {\bf 71}, 2625 (2002).
\bibitem{note} The oscillating behavior of the phase boundary (solid line) in Figs. \ref{fig7}(a) and (b) is due to discrete energy levels in a harmonic trap potential. To explain this in a simple manner, when we consider a two-dimensional harmonic potential $V({\bm r})=m\omega^2{\bm r}^2/2$, one particle energy is given by $E(n_x,n_y)=\omega[n_x+n_y+1]$ ($n_x,~n_y=0,1,2,\cdot\cdot\cdot$). In this case, the degeneracy of an eigen-energy $E=\omega[N_{xy}+1]$ equals $N_{xy}+1$. When degenerate energy levels below $E$ are fully occupied by $\uparrow$-spin atoms or $\downarrow$-spin atoms, the energy gap $\omega$ suppresses the superfluid phase transition to some extent, leading to the oscillation of the phase boundary in Figs. \ref{fig7}(a) and (b).
\bibitem{note2} The situation, $\mu_\uparrow\ne\mu_\downarrow$, actually occurs also in the case with trap-potential difference, as well as the case with mass imbalance. Figure \ref{fig7} indicates that a triplet pair amplitude is induced even when the difference of the chemical potentials only exists.
\bibitem{3loss1} J. Levinsen, N. R. Cooper, and V. Gurarie, Phys. Rev. Lett. \textbf{99}, 210402 (2007). 
\bibitem{3loss2} M. Jona-Lasinio, L. Pricoupenko, and Y. Castin, Phys. Rev. A \textbf{77}, 043611 (2008). 
\bibitem{3loss3} J. Levinsen, N. R. Cooper, and V. Gurarie, Phys. Rev. A \textbf{78}, 063616 (2008).
\bibitem{Gaebler1} J. P. Gaebler, J. T. Stewart, J. L. Bohn, and D. S. Jin, Phys. Rev. Lett. \textbf{98}, 200403 (2007).
\end{thebibliography}
\end{document}